\documentclass[%
 aip,
 amsmath,amssymb,
 reprint,%
]{revtex4-1}

\usepackage{graphicx}
\usepackage{dcolumn}
\usepackage{bm}

\usepackage[utf8]{inputenc}
\usepackage[T1]{fontenc}
\usepackage{mathptmx}
\usepackage{etoolbox}
\usepackage{physics} 
\usepackage{xcolor} 
\usepackage{soul} 
\usepackage{url} 
\usepackage{hyperref} 

\setstcolor{red}

\makeatletter
\def\@email#1#2{%
 \endgroup
 \patchcmd{\titleblock@produce}
  {\frontmatter@RRAPformat}
  {\frontmatter@RRAPformat{\produce@RRAP{*#1\href{mailto:#2}{#2}}}\frontmatter@RRAPformat}
  {}{}
}%
\makeatother
\begin{document}

\preprint{AIP/123-QED}

\title{Design of magnonic waveguides\\ using surface anisotropy-induced Bragg mirrors}
\author{Grzegorz Centa{\l}a}
 \email{grzcen@amu.edu.pl}
\author{Jaros{\l}aw W. K{\l}os}%
\affiliation{ 
Institute of Spintronics and Quantum Information, Faculty of Physics and Astronomy,               Adam Mickiewicz University,
Uniwersytetu~Pozna{\'n}skiego~2, 
61-614 Pozna{\'n},
Poland
}%

\date{\today}

\begin{abstract}
Waveguides are fundamental components for signal transmission in integrated wave-based processing systems. In this paper, we address the challenges in designing magnonic waveguides, including limitations such as non-uniform demagnetizing fields, reduced group velocity, and restricted operating frequency ranges. We propose a magnonic waveguide design with promising properties that overcome these limitations to a significant extent. Specifically, we investigate a waveguide formed within a uniform ferromagnetic layer ($\rm Co_{20}Fe_{60}B_{20}$) by applying surface anisotropy in strip regions, thereby creating Bragg mirror structures to confine spin waves and guide them along a single direction.
The proposed waveguide enables the propagation of high-frequency spin waves with high velocities in the ferromagnetic layer while minimizing static demagnetizing effects. We developed a model that allows for spin-wave confinement and guidance in two perpendicular directions by spatially modulating the surface anisotropy. The theoretical model was solved using the finite element method to calculate the dispersion relations of the waveguide modes and analyze their spatial profiles. Additionally, we determine the group velocity and localization characteristics, providing a comprehensive understanding of the waveguide’s performance.
\end{abstract}

\maketitle

Magnonics uses spin waves to process signals at gigahertz frequencies in nanoscale reconfigurable systems and is attractive compared to other wave-based computing technologies\cite{Wang_2021}, such as photonics or phononics, where nanoscale integration and reconfigurability are more challenging\cite{Li_2020,Wang_2024,Han_2024}. In magnonic systems\cite{Khitun_2004,Zheng_2022}, information is transmitted between functional blocks by waveguides, as in other wave-processing platforms.

\begin{figure}[b!]
\centering
  \includegraphics[width=1\columnwidth]{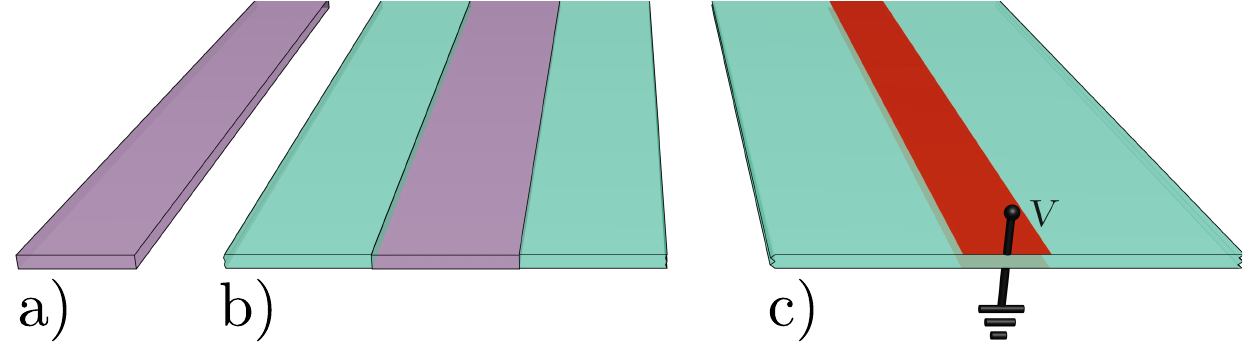}
  \caption{Selected designs of planar magnonic waveguides:
(a) Geometric confinement: a strip of ferromagnetic material (violet) embedded in a nonmagnetic matrix.
(b) Material parameter modification: a waveguide formed in the central region of a ferromagnetic layer by reducing the saturation magnetization (violet strip) compared to the surrounding ferromagnet (green areas).
(c) Voltage-controlled surface anisotropy: a waveguide induced by locally modified anisotropy (red region on the surface).}
  \label{fgr:BM_intro}
\end{figure}

Conventional magnonic waveguides are usually in the form of flat wires \cite{Kostylev_2007, Demidov_2008, Pirro_2011, Duerr_2012, Xing_2013} -- see Fig.~\ref{fgr:BM_intro}(a). However, their transmission properties are determined not only by geometry and material selection but also by the configuration of magnetization. In the absence of magnetocrystalline anisotropy and the lack of external magnetic fields, the shape anisotropy forces static magnetization to be oriented along the waveguide axis. In this configuration, the group velocity of the spin wave is significantly reduced \cite{Demidov_2008}. By applying the strong field, one can align the magnetization perpendicularly to the axis of the wire, which increases the group velocity; however, this simultaneity generates wells of static demagnetization field near the edges of the waveguide and induces so-called edge modes \cite{Duerr_2012,Xing_2013}.

Another technique to create a waveguide is the spatial modification of the material composition \cite{Obry_2013}, which is also used to fabricate planar one-dimensional magnonic crystals \cite{Ding_2013, Centala_2019}. If the saturation magnetization $M_{\rm s}$ is lower in the waveguide region -- see Fig.~\ref{fgr:BM_intro}(b), then we can confine and guide the spin waves of frequencies lower than the ferromagnetic resonance (FMR) frequency of the layer in which the waveguide is embedded can be confined. 

The most significant disadvantages of conventional waveguides, therefore, include: (i) limited reconfigurability - the geometry of the waveguide is fixed in the fabrication process, and changing its magnetic configuration must be done by applying a strong external magnetic field - which is technically difficult, (ii) the occurrence of strong static demagnetizing field (when the field is applied perpendicularly to the waveguide) or propagation speed limitation (when the field is applied along to the waveguide). A way to overcome these problems is to use a homogeneous layer as a conductor for spin waves with induced waveguide(s). 

Magnetic waveguides can be spontaneously formed in a uniform layer as magnetic domains \cite{albisetti_2016,hamalainen_control_2018,Wang_2018} or domain walls \cite{Garcia_2015,Wagner_2016} or imprinted by the temperature gradient \cite{vogel_2015}, inhomogeneous Oersted field \cite{Karenowska_2012}, inhomogeneous demagnetizing field \cite{Langer_2017}. One promising technique involves locally modifying the surface anisotropy to create magnonic waveguides within a uniform ferromagnetic layer -- see Fig.~\ref{fgr:BM_intro}(c). In such a system, the spin waves of lower frequencies can be guided within trenches of the effective magnetic field landscape shaped by the modification of the surface anisotropy \cite{Rana_2018,Li_2017}. This approach was also used to design various magnonic systems: magnonic crystals \cite{Choudhury_2020,Wang_2017}, tunnel junctions \cite{klos_hartman_2018}, circulators \cite{Zhao_2023}.

However, these solutions where the waveguide is imprinted in a ferromagnetic layer [Fig.\ref{fgr:BM_intro}(b,c)] allow for confinement and guidance of the spin waves only in a small range of frequencies limited up to the FMR frequency of the pristine layer.  Also, the strength of this confinement, i.e. the rate of exponential decay of the spin wave outside the waveguide, depends on the contrast of material parameters (saturation magnetization $M_{\rm s}$ or surface anisotropy constant $K_{s}$) and decreases to zero as the operating frequency approaches the FMR frequency of the uniform layer.

\begin{figure}[hb!]
\centering
  \includegraphics[width=1\columnwidth]{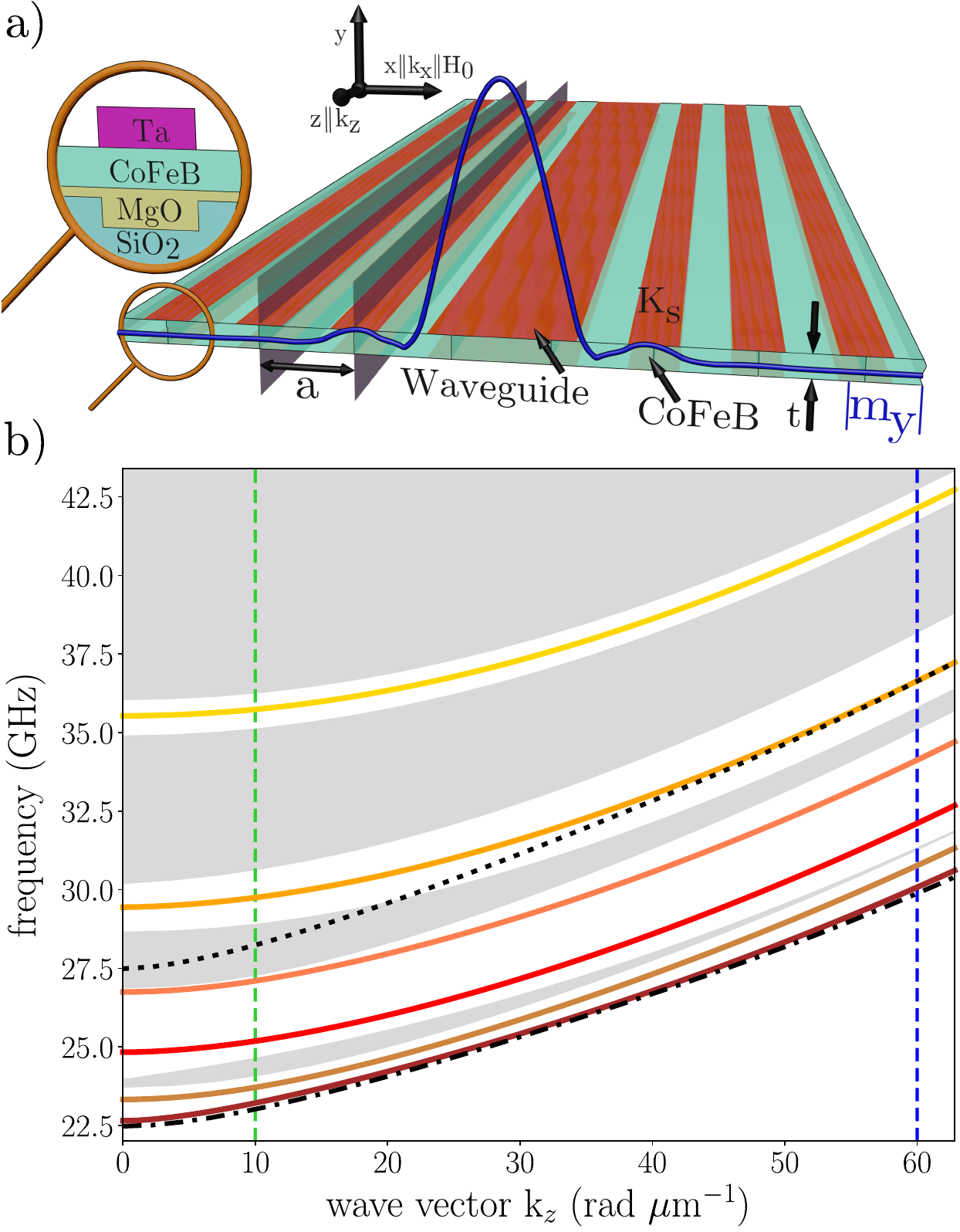}
  \caption{
(a) Structure under study: spin waves are confined by two Bragg mirrors (red stripes) formed by periodically modulated surface anisotropy on the top and bottom surfaces. Waves are guided along $z$ in the central region (wider strip -- defect). The anisotropy is set by $K_{\rm s}$ and the mirror period by $a$. The static field $H_0$ is applied in-plane, perpendicular to the waveguide. The blue line shows an exemplary out-of-plane profile $|m_y|$. The inset illustrates a possible experimental realization.
(b) Dispersion $f(k_z)$ of the waveguide modes (colored lines). Modes are confined when their frequencies fall into the band gaps of the magnonic crystals (white regions). The lower limit is the FMR of a uniform layer with continuous surface anisotropy on both sides (dash-dotted black line); for this design, higher-order modes (e.g., Nos. 5–6) can appear even above the FMR of a uniform layer without surface anisotropy (dotted line). Vertical dashed lines indicate the $k_z$ values used for mode profiles in Fig.~\ref{fgr:profiles}; for $k_z=10$~rad/$\mu$m we also evaluate a figure of merit for the trade-off between propagation and localization (see Supplemental~Material~2).}
\label{fgr:dispersion_relation}
\end{figure}

To overcome these limitations and realize the waveguide functionality for higher frequencies, we consider a pair of semi-infinite 1D magnonic crystals (acting as Bragg mirrors \cite{Morozova_2020}) to confine the spin waves in a linear defect \cite{Gallardo2018, Klos13}  -- see Fig.~\ref{fgr:dispersion_relation}(a). In the proposed system, spin waves can be confined in several frequency ranges, corresponding to the successive frequency gaps of the magnonic crystals. In our approach, the magnonic crystals will be created by applying surface anisotropy while preserving the uniformity of the magnetic layer \cite{Choudhury_2020}.
We will show that by appropriately playing with choosing the size of the structure, we can obtain both strong confinement of spin waves and a relatively high speed of their propagation along the waveguide.

We used  the Landau-Lifshitz equation to describe the magnetization dynamics in the considered system:
\begin{equation}\label{eq:LLE}
\frac{d\textbf{M}}{dt}=-\gamma\mu_{0}\Big( \textbf{M} \times \textbf{H}_{\rm eff} + \frac{\alpha}{M_{\rm s}} \textbf{M} \times (\textbf{M} \times \textbf{H}_{\rm eff})\Big),
\end{equation}

where $\textbf{M}(\textbf{r},t)$ is magnetization, $\gamma$ is gyromagnetic ratio, $\mu_{0}$ is vacuum permeability, $\alpha$ is damping constant, and $\textbf{H}_{\rm eff}(\textbf{r},t)$ is effective field. Damping is neglected for the considered low-damping material.

For the in-plane applied field, the sample is magnetically saturated. As a result, we can easily linearize Eq.~\ref{eq:LLE} by taking  $\mathbf{M}(\textbf{r},t)~=~M_{\rm s}\hat{\mathbf{x}}_M~+~\mathbf{m}(\mathbf{r})e^{i\omega t}$, where the static component is oriented along $\hat{\mathbf{x}}_M$ direction and dynamic component oscillates harmonically in time \cite{Gurevich_1996}.

The effective magnetic field $\textbf{H}_{\rm eff}(\textbf{r},t)$ consists of the external field of $H_{0}\hat{\mathbf{x}}$, the exchange field of $\textbf{H}_{\rm ex}(\textbf{r},t)= \left( 2A/\mu_{0} M^2_{\rm s} \right)\Delta\textbf{M}(\textbf{r},t)$, and the dipolar 
field of $\textbf{H}_{\rm d}(\textbf{r},t)=-\nabla \varphi (\mathbf{r},t)$. The symbols $A$ and $M_{\rm s}$ stand for exchange stiffness constant and saturation magnetization, respectively. The magnetostatic potential ($\varphi$) is determined using the Gauss law for magnetism, under the magnetostatic approximation $\Delta\varphi=\div{\textbf{m}}\;$ \cite{Szulc_2022}.

Uniaxial surface anisotropy is introduced by boundary conditions. 
We took a general form for the boundary conditions for spin waves in the presence of uniaxial out-of-plane anisotropy\cite{Gurevich_1996, Centala_2019, Centala_2023}:
\begin{equation}\label{eq:BC}
\hat{\mathbf{x}}_M\times\frac{\partial \mathbf{m}}{\partial \hat{\mathbf{n}}}+\frac{K_{\rm s}}{A}\Big(\left(\hat{\mathbf{n}}\cdot\mathbf{m}\right)\hat{\mathbf{n}}\times\hat{\mathbf{x}}_M+\left(\hat{\mathbf{n}}\cdot\hat{\mathbf{x}}_M\right)\hat{\mathbf{n}}\times\mathbf{m}\Big)=0,
\end{equation}

where $\hat{\mathbf{n}}$ is surface normal. For $\hat{\mathbf{x}}_M=\hat{\mathbf{x}}$ and $\hat{\mathbf{n}}=\hat{\mathbf{y}}$ [Fig.~\ref{fgr:dispersion_relation}(a)], expression (\ref{eq:BC}) simplifies to:

\begin{equation}\label{eq:BCsimp}
    \begin{split}
        \frac{\partial m_{y}}{\partial n} - \frac{K_{s}}{A}m_{y}=0,
        \\
        \frac{\partial m_{z}}{\partial n}=0.
    \end{split}
\end{equation}
It is worth noting that the surface anisotropy constant $K_{\rm s}$ is the only spatially dependent parameter in the system. Therefore, the patterning of the Bragg mirrors for spin waves is achieved by spatially varying the boundary conditions, due to the $K_{\rm s}(x)$ dependence.

The linearized form of Eq.~\ref{eq:LLE} takes the form:

\begin{equation}\label{eq:linear_LLE}
    \begin{split}
        -i\omega m_{y}+\frac{2A\gamma}{M_{s}}\Delta m_{z}-\frac{2A\gamma}{M_{s}}k^{2}_{z}m_{z}-\gamma \mu_{0} H_{0} m_{z}-i M_{s} \gamma \mu_{0} k_{z} \varphi=0, \\
        -i\omega m_{z}-\frac{2A\gamma}{M_{s}}\Delta m_{y}+\frac{2A\gamma}{M_{s}}k^{2}_{z}m_{y}+\gamma \mu_{0} H_{0} m_{y}+M_{s} \gamma \mu_{0} \pdv{\varphi}{y}=0
    \end{split}
\end{equation}
along with the Gauss equation for magnetism inside ferromagnet ($\textbf{m}\neq 0$) and outside ferromagnet ($\textbf{m} = 0$):
\begin{equation}\label{eq:gauss_ferromagnet}
        \pdv[2]{\varphi}{x}+\pdv[2]{\varphi}{y}-k^2_{z}\varphi-\pdv{m_{y}}{y}-i k_zm_{z}=0.
\end{equation}
For uniformity in $z-$direction, solutions take plane wave form $e^{i k_z z}$ with amplitudes $m_y(x,y)$, $m_z(x,y)$, $\varphi(x,y)$. Eqs.~\ref{eq:linear_LLE},~\ref{eq:gauss_ferromagnet} together with boundary conditions (\ref{eq:BC}) were implemented in the mathematical module of the COMSOL Multiphysics\cite{Szulc_2022} and solved via finite element method for:

\noindent
(i) \textit{Unit cell of infinite magnonic crystal} – determining frequency gaps: Bloch boundary conditions in $x$-direction for 2D wave vector $\mathbf{k} = k_{x} \hat{\mathbf{x}} + k_{z} \hat{\mathbf{z}}$ were applied to find gap edges of magnonic crystal for given value of $k_z$.

\noindent
(ii) \textit{Full waveguide structure} – determining the 1D dispersion of waveguide modes:  periodic boundary conditions were applied at both ends of the whole structure (two 21-period Bragg mirrors) to enable the study of weakly localized waveguide modes.

Both systems include non-magnetic regions ($1000~\times$ thickness) above and below considered structure to correctly calculate dynamic demagnetizing field effects.

We investigated the structure shown in Fig.~\ref{fgr:dispersion_relation}(a): a 6-nm-thick Co${_{20}}$Fe${_{60}}$B${_{20}}$ layer with low spin-wave damping ($\alpha = 0.0029$) \cite{Wang_2022}. Surface anisotropy was applied at both Co${_{20}}$Fe${_{60}}$B${_{20}}$ interfaces. The regions with surface anisotropy (red strips) are arranged periodically, forming Bragg mirrors that confine spin waves in a wider central strip. The structure thus acts as a defected magnonic crystal with a 100-nm period (50-nm anisotropy strips alternating with 50-nm wide sections of pristine Co$_{20}$Fe$_{60}$B$_{20}$ layer) and a defect region defined by a single 150-nm-wide anisotropy strip. Volume magnetocrystalline anisotropy was neglected.

We used experimental values: saturation magnetization $M_{\rm s}$~=~1150~kA/m\;\cite{Lee_2014}, exchange stiffness $A$~=~28~pJ/m\;\cite{Wang_2022}, and gyromagnetic ratio $|\gamma|$ = 176~rad/T/ns\;\cite{Wang_2022}. 
We assumed a surface anisotropy $K_{\rm s}=1.05$~mJ/m$^{2}$ at both interfaces, corresponding to an effective volume anisotropy $K_{\rm v}=2K_{\rm s}/t$ for CoFeB/MgO with a Ta cap~\cite{Lee_2014}; see the inset in Fig.~\ref{fgr:dispersion_relation}(a). Experiments~\cite{Lee_2014} do not separate the CoFeB/MgO and CoFeB/Ta contributions, but for a thin ferromagnetic layer this is not critical: Supplementary Material~1 shows that using $K_{\rm s}=2.1$~mJ/m$^{2}$ at a single interface yields a dispersion almost identical to that in Fig.~\ref{fgr:dispersion_relation}(b).
Since the CoFeB/MgO interfacial anisotropy nearly vanishes for ultrathin MgO\cite{Yamanouchi_2011}, a structure can be fabricated by modulating the MgO thickness. The structure can be manufactured by (i) patterning trenches in SiO$_2$, (ii) sputtering MgO ($<0.5$~nm on ridges, 1~nm in trenches), (iii) depositing a 6~nm CoFeB layer and Ta cap, and (iv) selectively removing Ta above the SiO$_2$ ridges using focused ion-beam milling.

An external field ($\mu_0H_0$ = 500~mT) was applied in the $x$-direction, perpendicular to the stripes and tangential to the film. In this Damon-Eshbach geometry, spin waves propagate perpendicular to the field, resulting in high group velocity. Unlike the ferromagnetic strip [Fig.~\ref{fgr:BM_intro}(a)], this structure avoids edge modes due to the absence of static demagnetization.

The choice of a 6 nm thickness represents a balance between high group velocity, which increases with $t$, and strong effective anisotropy, which decreases with $t$ (see Supplementary~Material~2). A large anisotropy contrast allows for wide frequency gaps and strong confinement, while a 50\% filling fraction further enhances the gap width. The 100 nm lattice constant provides a broad first Brillouin zone, giving access to multiple low-frequency bands at comparatively high frequencies.

Due to magnetic anisotropy, the ferromagnetic resonance (FMR) frequency in a ferromagnetic layer is non-zero for non-zero applied field. Above the FMR frequency, the wave vector becomes real, enabling spin waves to propagate. Below the FMR frequency, it becomes complex, resulting in unphysical exponential changes in spin-wave amplitude. This can be exploited to build a waveguide (see Fig.~\ref{fgr:BM_intro}(c)). By tuning material parameters in the waveguide region, spin-wave profiles become oscillatory inside and decay exponentially in the surrounding layer, but only for frequencies up to the FMR. This limit is lifted using a pair of magnonic crystals as Bragg mirrors. The system (see Fig.~\ref{fgr:dispersion_relation}(a)) overcomes this constraint and enables waveguiding at higher frequencies, within frequency gaps where complex wave vectors ensure confinement.

\begin{figure}[t]
\centering
  \includegraphics[width=1\columnwidth]{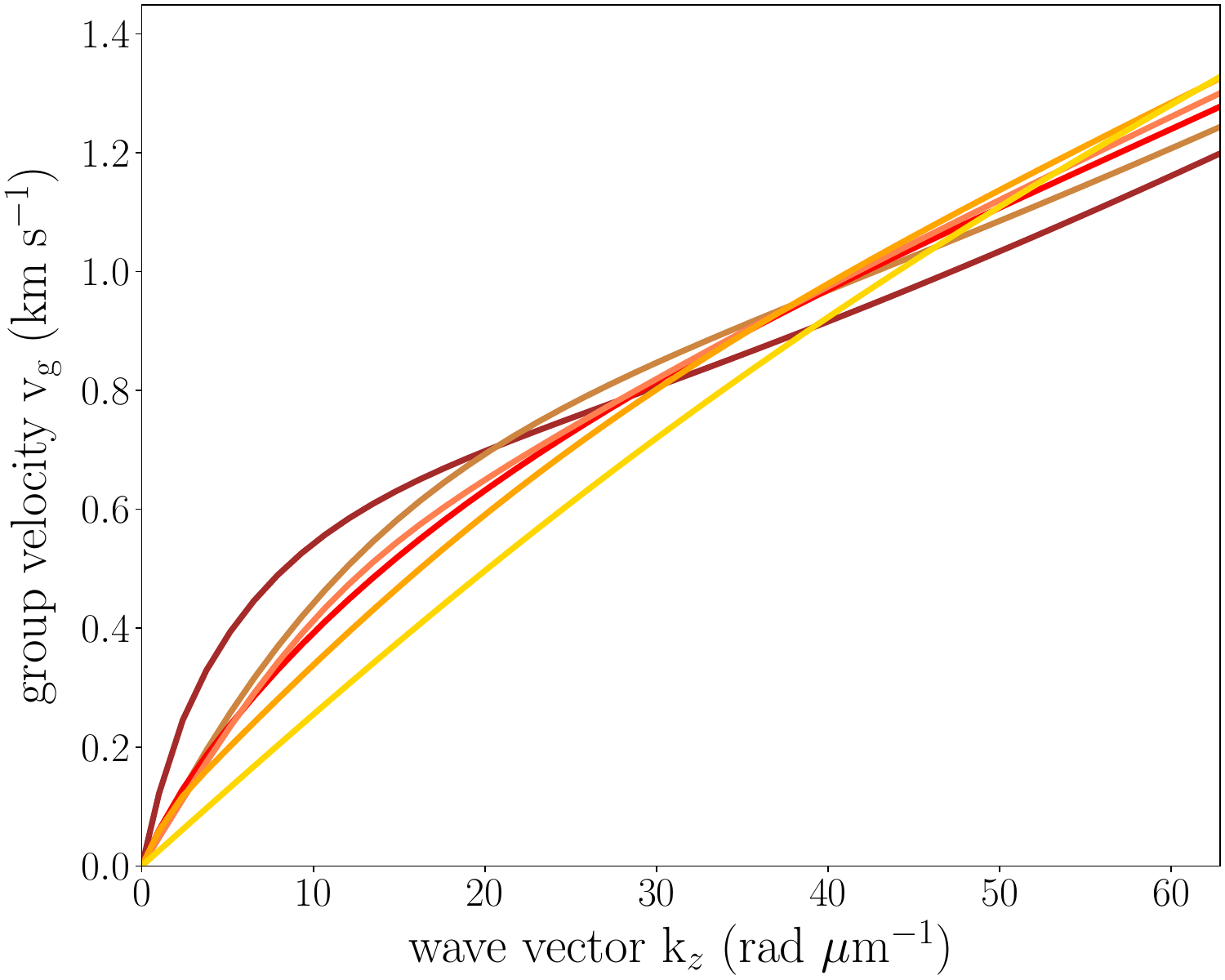}
  \caption{Group velocity of the waveguide modes as a function of the wave vector $v_{g}(k_{z})$. Line colors denote successive modes with increasing frequencies (see Fig.~\ref{fgr:dispersion_relation}(b)).
  }
  \label{fgr:SW_velocity}
\end{figure}

This feature is clearly visible in Fig.~\ref{fgr:dispersion_relation}(b), where the waveguide modes (solid lines) appear in several frequency gaps (white regions), including those located well above the FMR frequency of the uniform layer (dotted black line). These modes propagate along the waveguide with the real wave number $k_z$ and decaying when penetrating the Bragg mirrors at the exponential rate given by the imaginary part of the wave vector $k_x$ from the frequency gap of the magnonic crystals (Bragg mirrors).
Within the considered frequency range (20–45 GHz), one or two waveguide modes can be observed in each successive frequency gap. However, the number of these modes can increase in each gap if a wider waveguide is considered. The number of waveguide modes in a given gap also depends on the width of the gap, which, in turn, is determined by the parameters of the Bragg mirrors.

Let us discuss in more detail the advantages of the proposed design. The large group velocity and strong localization are essential features of the waveguide that determine its performance. Only a sufficiently high value of the group velocity allows for the transmission of spin-wave signals over practical distances in the presence of damping. On the other hand, strong localization within the waveguide reduces cross-talk between different waveguides in the system.

Fig.~\ref{fgr:SW_velocity} shows the group velocity of the waveguide modes, $v_{g} = d\omega/dk_{z}$, calculated from the dispersion relation. The group velocity of all modes increases with the wave vector $k_z$. For smaller values of $k_z$, the waveguide modes with the lowest frequencies (brownish and reddish lines) exhibit higher group velocities; however, for larger $k_z$, this order is reversed. The obtained values of group velocity are suitable for spin-wave transmission in low-damping materials such as CoFeB. However, the transmission of spin waves with long wavelengths is inefficient, as $v_{g}$ approaches zero for $k_z \rightarrow 0$. Knowing the group velocity, we can calculate the spin-wave decay length $L_d$ in the presence of Gilbert damping, as discussed in Supplementary Material, Section~3, where we show that for CoFeB, $L_d$ defined by $m_y(z+L_d)/m_y(z)=1/e$ is significantly larger than a few waveguide (defect) widths.

The localization of waveguide modes is primarily determined by the complex wave vector, $k_{x}$, within the frequency gaps of the magnonic crystals. It is well established \cite{Kohn_1959} that the imaginary part of the wave vector reaches its maximum within the frequency gap, indicating the strongest localization, and decreases toward the band edges. However, we used a general measure of localization: the Inverse Participation Ratio (IPR), which can be calculated from the profile of waveguide modes, $|m_y|$, across the entire structure. We computed a discrete approximation of the IPR\cite{Thouless_1974}:
\begin{equation}\label{eq:IPR}
\widetilde{\rm IPR}={\rm IPR}\frac{S}{m n}=  \frac{ \sum_{i,j}^{m,n} |m_{y}(x_i,y_j)|^{4}  }{ \left( \sum_{i,j}^{m,n} |m_{y}(x_i,y_j)|^{2} \right)^2 },
\end{equation}
with $S=43at$ - cross-section area of the ferromagnetic layer, $n=36$ - points across the thickness, $m=4300$ - points along the layer. $\widetilde{\rm IPR}$ ranges from $1/(mn)$ (uniform distribution) to 1 (point localization).

\begin{figure}[t]
\centering
  \includegraphics[width=1\columnwidth]{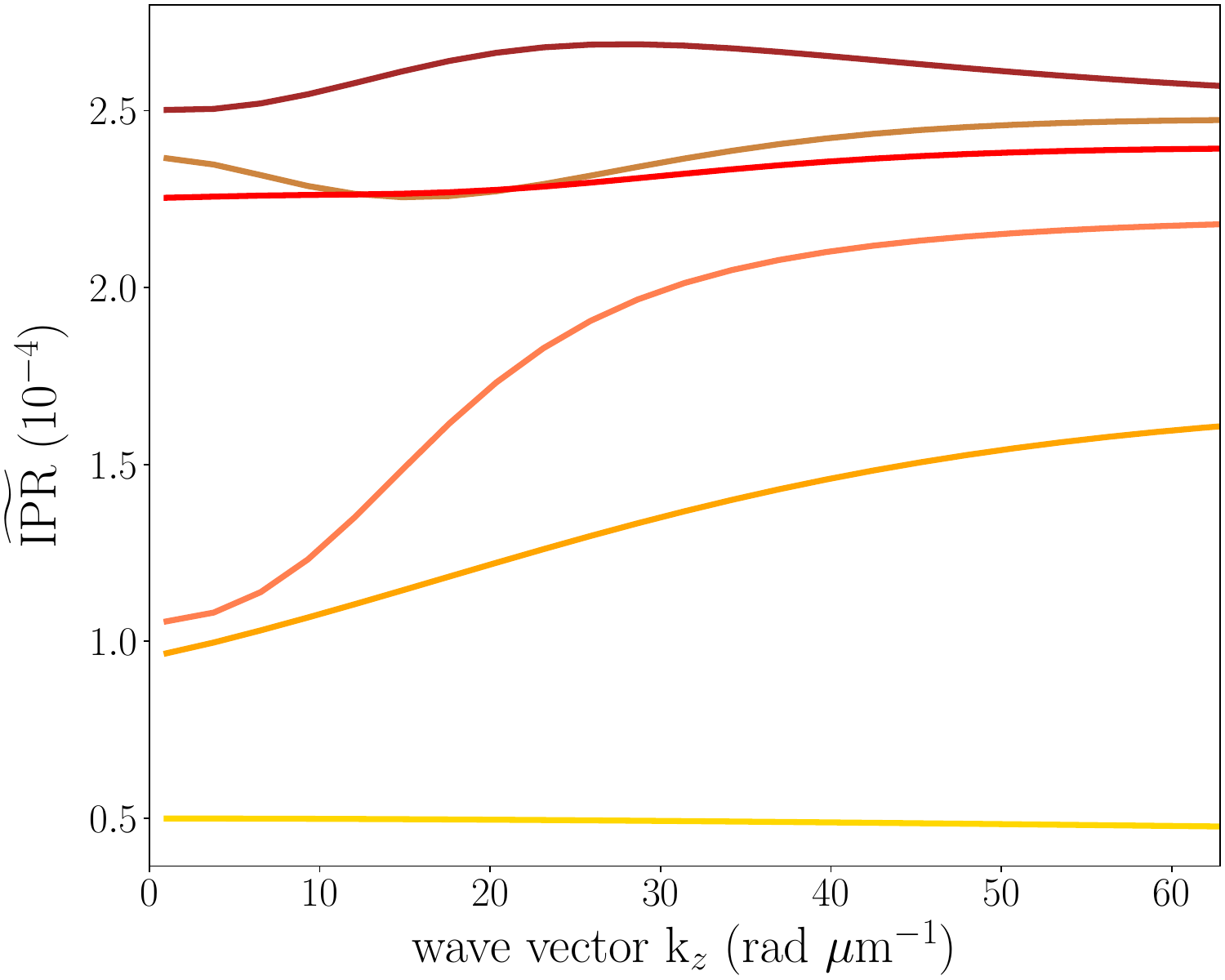}
  \caption{Measure of the localization of the waveguide modes: modified inverse participation ratio  as a function of the wave vector $\widetilde{\rm IPR}(k_{z})$ -- see Eq.\ref{eq:IPR}. Line colors denote successive modes with increasing frequencies (see Fig.~\ref{fgr:dispersion_relation}(b)).} 
  \label{fgr:IPR}
\end{figure}

\begin{figure}[ht!]
\centering
  \includegraphics[width=1\columnwidth]{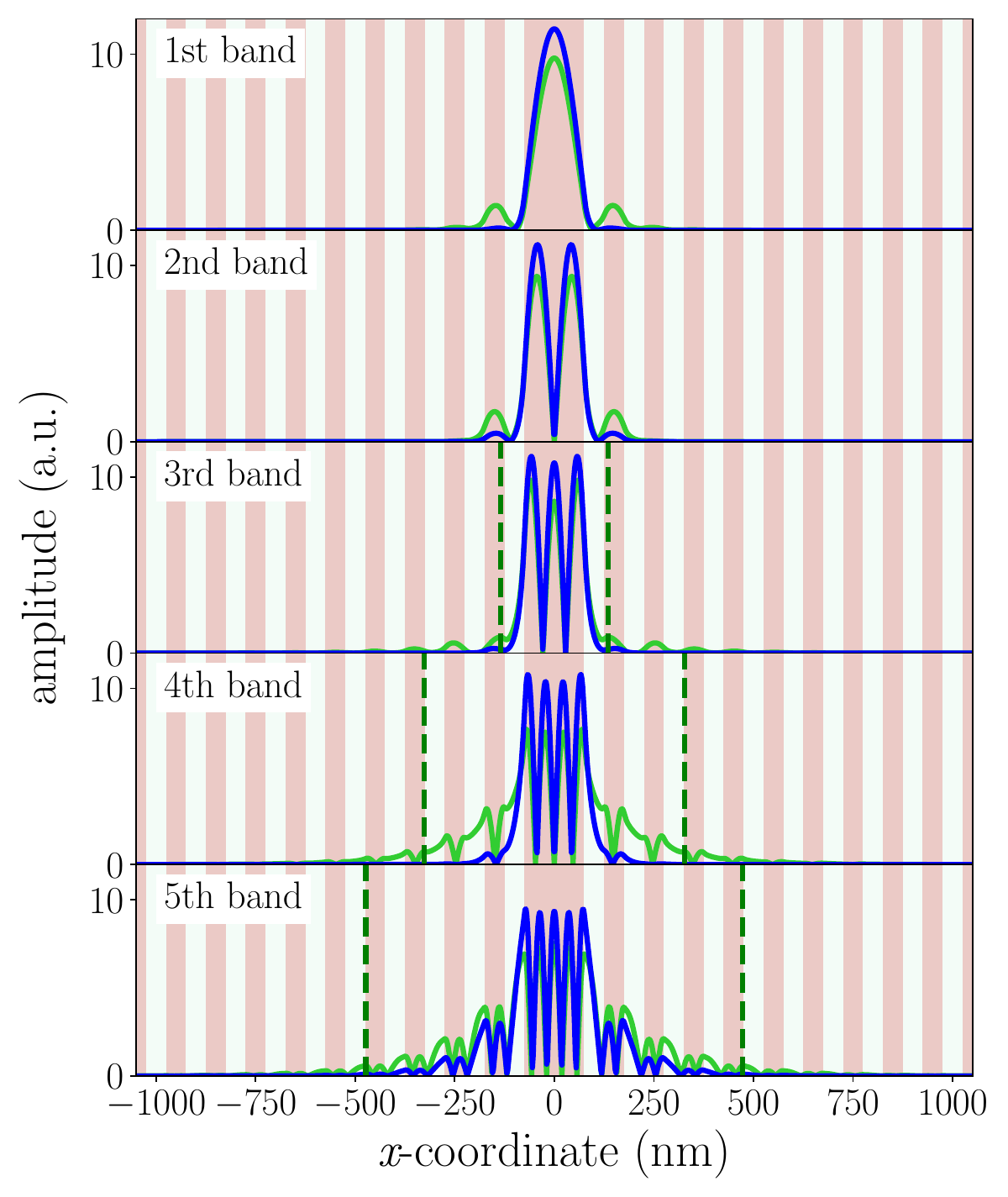}
  \caption{Profiles of the out-of-plane component of the dynamic magnetization, $|m_y|$, for successive waveguide modes at two selected wave vectors: $k_{z}=10$ rad/$\mu$m (green) and $k_{z}=60$ rad/$\mu$m (blue). See Fig.~\ref{fgr:dispersion_relation}(b) for the corresponding dispersion branches. Pink (white) regions indicate sections of the ferromagnetic layer with (without) surface anisotropy. Green dashed lines indicate the isolation distance $L_i$ necessary to prevent crosstalk between adjacent waveguides in integrated circuits.
  }
  \label{fgr:profiles}
\end{figure}

Fig.~\ref{fgr:IPR} shows the dependence of $\widetilde{\rm IPR}$ on the wave vector $k_z$ for successive waveguide modes. The values are highest for modes 1–3 (brownish and reddish lines), as reflected in the mode profiles in Fig.~\ref{fgr:profiles}. These modes have frequencies below the FMR frequency of the uniform ferromagnetic layer, meaning the spin wave must tunnel through regions of the Bragg mirror without surface anisotropy. This results in faster spatial decay across the mirrors: $\pm e^{\mp\Im(k_x) x}$. Such decay corresponds to wider frequency gaps, where $\Im(k_x)$ is larger.

Higher-frequency modes (e.g., Nos.~4–6 -- yellowish and pink lines) can only be localized by Bragg mirrors. For modes like No.~4 and No.~5, localization strength varies with $k_z$. This is seen in the frequency shifts of these modes in Fig.~\ref{fgr:dispersion_relation}(b). As $k_z$ increases, mode No.~4 moves away from the gap edge, while No.~5 approaches it. Both a mode’s proximity to the gap center and the gap’s width are known \cite{Kohn_1959} to enhance localization in periodic structures via increased $\Im(k_x)$. These effects are visible in the profiles of modes No.~4 and No.~5 in Fig.~\ref{fgr:profiles} (green and blue lines correspond to different $k_z$ values).

Notably, $\widetilde{\rm IPR}$ and $v_g$ are interdependent. Increasing the ferromagnetic layer thickness $t$ boosts $v_{g}$\cite{Stancil_2009}, but reduces the effective anisotropy $K_{\rm eff} = -\mu_{0}M^{2}_{\rm s} + K_{\rm s}/t$, weakening the spatial modulation. This reduces the Bragg mirror effect, narrows the frequency gaps, and lowers localization strength (as measured by $\widetilde{\rm IPR}$). We discuss the optimal layer thickness and the balance between $v_{g}$ and $\widetilde{\rm IPR}$ in Supplemental~Material~2.

The system is a multimode waveguide where successive modes differ in the number of nodes [Fig.~\ref{fgr:dispersion_relation}(b)]. Mode profiles (blue and green lines in Fig.~\ref{fgr:profiles}) reveal two distinct regions: the defect (wider red strip in Fig.~\ref{fgr:profiles}) with nearly constant amplitude, and the Bragg mirrors with exponentially decaying amplitude. The number of nodes increases from 0 to 4 (for modes 1–5) with frequency, regardless of whether the modes appear singly or in pairs within the gaps [Fig.~\ref{fgr:dispersion_relation}(b)].
The IPR is related to the distance ensuring isolation between two waveguides, $L_i$, which can be determined from the SW profile using the condition
$10 \log_{10}(|m_{y}(L_i)|^2/|m_{y,0}|^2) = -20$~dB, where $|m_{y,0}|$ is the maximum dynamic magnetization inside the waveguide. For $k = 10$~rad/$\mu$m (green lines in Fig.~\ref{fgr:profiles}), this yields $L_i$ = 0.40, 1.68, 2.66, and 6.23 defect widths for modes 3--6; for the more focused modes 1 and 2, $L_i$ is even smaller but difficult to extract reliably from $m_y(x)$.

The proposed design offers several advantages: it uses a homogeneously magnetized, low-damping ferromagnetic film and requires only periodic patterning of a nonmagnetic over(under)layer to form anisotropy-induced Bragg mirrors, suppressing static demagnetizing fields and eliminating edge modes typical of strip waveguides. Bragg confinement also enables guiding of spin waves at frequencies above the ferromagnetic resonance of a uniform layer, which simpler designs cannot achieve. This, however, involves a trade-off between high group velocity and strong confinement, requiring relatively thin ferromagnetic layers and sufficiently large surface anisotropy.

\section*{Supplementary Material}

The supplementary material is divided into three sections. The first section presents the dispersion relation for a doubled surface anisotropy constant applied at a single surface (e.g., the top surface only), as well as the dispersion relation obtained using realistic values for voltage-controlled magnetic anisotropy. The second section provides the figure of merit calculated for our structure. The third section describes the methodology for calculating the decay length and presents its dependence on the wave vector.

\begin{acknowledgments}
This work has received funding from National Science Centre Poland grants  UMO-2020/39/O/ST5/02110, UMO-2021/43/I/ST3/00550, and support from the Polish Science National Agency for Academic Exchange grant BPN/PRE/2022/1/00014/U/00001.
\end{acknowledgments}

\section*{Data Availability Statement}

The data for the essential figures (Fig.~\ref{fgr:dispersion_relation}(b), Fig.~\ref{fgr:SW_velocity}, Fig.~\ref{fgr:IPR}) can be accessed via the following https://doi.org/10.5281/zenodo.14580873. The remaining data is available from the corresponding author on reasonable request.

\section*{Conflict of Interest Statement} 

The authors have no conflicts to disclose.

\section*{Author contributions statement}
G. C.: Data curation, Formal analysis, Funding acquisition, Investigation, Methodology, Software, Validation, Visualization, Writing – original draft, Writing – review and editing. J. W. K.: Conceptualization, Data curation, Formal analysis, Funding acquisition, Investigation, Methodology, Project administration, Resources, Software, Supervision, Validation, Visualization, Writing – original draft, Writing – review and editing.

\section*{References}

%

\end{document}